\documentclass[fleqn,usenatbib]{mnras}

\usepackage{newtxtext,newtxmath}
\usepackage[T1]{fontenc}

\DeclareRobustCommand{\VAN}[3]{#2}
\let\VANthebibliography\thebibliography
\def\thebibliography{\DeclareRobustCommand{\VAN}[3]{##3}\VANthebibliography}

\usepackage{graphicx}	% Including figure files
\usepackage{amsmath}	% Advanced maths commands

\title[Constraining DM Profiles with Wide Binaries]{Constraining Dark Matter Density Profiles in UFDs with Wide Binaries: Forecast for the Chinese Space Station Survey Telescope}

\author[Y. Tao et al.]{
Yixi Tao$^{1}$, 
Haijun Tian$^{2}$ \thanks{E-mail: hjtian@hdu.edu.cn}, 
Bin Yue$^{3}$,
and Jorge Pe\~narrubia$^{4}$ \thanks{E-mail: jorpega@roe.ac.uk}
\\
$^{1}$School of Science, Hangzhou Dianzi University, Hangzhou 310018, China.\\
$^{2}$National Astronomical Data Center Zhejiang Branch, Hangzhou 310018, China.\\
$^{3}$National Astronomical Observatories, Chinese Academy of Sciences, 20A Datun Road, Chaoyang District, Beijing, 100101, China\\
$^{4}$Institute for Astronomy, University of Edinburgh, Royal Observatory, Blackford Hill, Edinburgh EH9 3HJ, UK. \\
}

\date{Accepted XXX. Received YYY; in original form ZZZ}

\pubyear{\the\year{2025}}

\begin{document}
\label{firstpage}
\pagerange{\pageref{firstpage}--\pageref{lastpage}}
\maketitle

% Abstract of the paper
\begin{abstract}
{
 The internal structure of dark matter halos on sub-galactic scales remains a key open question, particularly in the context of the core–cusp problem. Ultra-faint dwarf galaxies (UFDs), owing to their extreme dark matter dominance, provide a promising laboratory to probe these density profiles through stellar tracers.
In this work, we assess the capability of the Chinese Space Station Telescope (CSST) to detect and characterize wide binary stars in the nearby UFD Segue\,I, using mock observations.We generate mock binary populations based on our existing $N$-body simulations and incorporate realistic CSST observational conditions, including the expected deep-field limiting magnitude ($g \sim 27.5$ mag) and a photometric completeness of approximately $90\%$. The two-point correlation function (2PCF) of stellar pairs is used as a statistical tool to recover the binary fraction under these assumptions.
We find that CSST can robustly detect wide binaries at the $3\sigma$ level for binary fractions as low as $f_b \gtrsim 0.01$, provided a stellar sample size of $N_{\rm star} \gtrsim 2300$. However, distinguishing between cusped and cored dark matter profiles is significantly more demanding, requiring $N_{\rm star} \gtrsim 6000$ and $f_b \gtrsim 0.1$  within $\sim 40$~kpc.
}
\end{abstract}


%%%%%%%%%%%%%%%%%%%%%%%%%%%%%%%%%%%%%%%%%%%%%%%%%%

%%%%%%%%%%%%%%%%% BODY OF PAPER %%%%%%%%%%%%%%%%%%

\section{Introduction}
In the mass--energy budget of the Universe, dark matter (DM), which contributes about 26.8\%, remains one of the most profound mysteries in modern physics. %The existence of this invisible component was first proposed by \citep{Zwicky1933} through the study of the dynamics of the Coma Cluster.
The existence of this invisible component was not first proposed by a single author; rather, early dynamical estimates date back to the late 19th century \citep{Bertone2018}. A seminal contribution, however, came from \citet{zwicky1933}, who inferred the presence of what he called "dark matter" from the dynamics of the Coma Cluster.Since then, multiple independent lines of evidence—including gravitational lensing \citep{Tyson1990,Clowe2006}, the anisotropies of the cosmic microwave background \citep[CMB,][]{Planck2018VI,Spergel2003}, and large-scale structure formation reproduced by cosmological N-body simulations  \citep{Springel2005,BoylanKolchin2009}—have established dark matter as a fundamental component of the Universe.Within the standard cold dark matter (CDM) cosmological paradigm, high-precision $N$-body simulations have successfully predicted the hierarchical assembly of structure on scales ranging from the cosmic web to galaxy clusters and galaxies. However, on galactic scales, CDM faces several long-standing challenges. One of the most prominent is the so-called ‘core or cusp problem’,which refers to the discrepancy between the cuspy central density profiles \citep[e.g., Navarro--Frenk--White, NFW;][]{Navarro1996} predicted by CDM-only simulations and the cored, constant-density profiles inferred from observations of dwarf galaxies \citep{Moore1994,deBlok2001,Oh2011,Read2019}. Understanding whether low-mass dark matter halos truly possess cores or cusps provides a crucial test of CDM on small scales and may even hint at the nature of DM itself \citep[e.g., self-interacting DM:][]{Spergel2000,Fitts2017}.

It is important to note that, while baryonic feedback processes can significantly modify the inner dark matter density profiles of more massive dwarf galaxies, their impact is expected to be minimal in UFDs. Due to their extremely low stellar masses and inefficient star formation, these systems likely lack sufficient energetic feedback to induce substantial dark matter heating. As a result, several studies suggest that UFDs are expected to retain their primordial cuspy density profiles \citep{Penarrubia2012}, consistent with extrapolations of the stellar-to-halo mass dependence found in simulations of more massive systems \citep{DiCintio2014}. However, recent work indicates that stochastic processes—such as merger-driven heating and fluctuations in the gravitational potential—can still drive core formation or introduce significant scatter in the central density slopes \citep{Orkney2021,Orkney2022}. This makes UFDs a particularly compelling environment in which to test the core or cusp problem.

Ultra-faint dwarf galaxies (UFDs) offer unique laboratories for probing the distribution of dark matter (DM) on the smallest accessible scales. As the faintest and most DM-dominated galaxies known \citep{Strigari2008,Simon2019}, UFDs often exhibit mass-to-light ratios exceeding several hundred. Consequently, the stellar kinematics within these systems are almost entirely governed by the gravitational potential of their DM haloes, making them ideal targets for constraining the inner density profiles of DM. Segue\,I, located at a heliocentric distance of $\sim 23$ kpc \citep{Belokurov2007b,Simon2011}, is a well-studied ultra-faint dwarf spheroidal galaxies (dSphs) often prioritized for such dynamic analyses. However, this very characteristic of DM dominance is tied to an extreme sensitivity to observational artifacts: its extremely low intrinsic velocity dispersion ($\sigma_{\mathrm{int}} \sim 1$--$4~\mathrm{km~s^{-1}}$) implies that even modest systematic uncertainties can dramatically bias dynamical mass estimates. A chief concern among these systematics is unresolved binary stars. The orbital motions of binaries introduce spurious line-of-sight velocity variations which artificially inflate the observed velocity dispersion ($\sigma_{\mathrm{obs}}^2 = \sigma_{\mathrm{DM}}^2 + \sigma_{\mathrm{bin}}^2$). Given the minimal $\sigma_{\mathrm{DM}}$ in UFDs, \citet{McConnachie2010} showed that unresolved binaries could boost $\sigma_{\mathrm{obs}}$ to values consistent with observation, potentially mimicking a significantly higher DM content. The challenge is further complicated by the high binary fractions \cite[e.g., $\sim 42$--$44\%$ in classic UFDs;][]{Spencer2018} and a preference for long orbital periods, necessitating extensive multi-epoch data to obtain pure DM-driven kinematic measurements.

The binary populations of dwarf galaxies are both ubiquitous and diverse. Their properties---such as binary fraction and orbital parameter distributions---may encode information about the star formation history, chemical enrichment, and dynamical evolution of their host galaxies \citep{Minor2013, Martinez2011,Koposov2011}. Wide binaries (separations $s \gtrsim 1000$ au) are particularly interesting. Their demographics are initially set by star formation processes but are subsequently modified by dynamical perturbations. Due to their large cross sections and low binding energies, wide binaries are extremely susceptible to gravitational perturbations from field stars, molecular clouds, and DM subhalos \citep{Chandrasekhar1944,Heggie1975,Bahcall1985,Weinberg1986,Jiang2010}. Consequently, they are sensitive probes of the small-scale structure of the gravitational potential. The prospect of using wide binaries as dynamical tracers is particularly compelling in UFDs: unlike larger galaxies, UFDs are expected to retain steep central cusps due to their shallow potential wells and inefficient baryonic feedback \citep{Wheeler2015,Fitts2017}. Moreover, their ancient stellar populations \citep[formed $\sim 13$--$14$ Gyr ago;][]{Brown2014} have had ample time to interact with substructures, allowing their present-day binary population -- or its depletion---to directly constrain both the survival of DM cusps and the abundance of DM subhalos \citep[][hereafter, P16]{Penarrubia2016}.

The formation of wide binaries remains an open question. Traditional star formation scenarios struggle to explain how pairs with separations $\gtrsim 10^3$~au can form and survive in the dense environments of embedded clusters, where tidal interactions and close encounters efficiently disrupt soft systems. Despite these challenges, several mechanisms have been proposed to account for their existence. These include the fragmentation of turbulent gas cores \citep{Lee2019} and the dynamical unfolding of compact triple systems, which can eject a distant companion \citep{Reipurth2012}. A comprehensive review of the various pathways to wide binary formation is provided by \citet{Duchene2013}. However, recent theoretical work has demonstrated that wide binaries can also arise naturally through dynamical processes following cluster dissolution. \citet{Penarrubia2020} showed that in tidal streams produced by cluster disruption, unrelated stars can become gravitationally bound through chance low-velocity interactions, with the formation efficiency scaling with the local phase-space density of the stream. This mechanism allows wide binaries to form not only during the early expansion of star-forming regions but also long after cluster dispersal, providing a pathway for their creation even in low-density galactic environments. Observationally, \citet{Shariat2025} found that UFDs such as Bo\"otes\,I do host wide binaries with separations of several thousand astronomical units, with demographics similar to those in the Solar neighborhood \citep{tian2020,ElBadry2021}, suggesting that their formation is not strongly suppressed in metal-poor, low-mass galaxies. Once formed, however, their subsequent evolution in UFDs can be strongly affected by the surrounding dark matter potential.

However, detecting and characterizing wide binaries in UFDs is challenging. Their large distances and sparse, faint stellar populations make it difficult to distinguish physical pairs from chance projections, especially in the absence of high-precision parallaxes and proper motions. Hubble Space Telescope (HST) imaging has offered tentative hints of wide binaries in some UFDs but suffers from low statistical significance and contamination uncertainties. Therefore, next-generation facilities are essential.

Despite these advances, our understanding of the formation and primordial properties of wide binaries remains highly incomplete. The relative importance of the proposed formation channels, the initial binary fraction, and the distributions of orbital parameters in low-mass, metal-poor systems such as ultra-faint dwarf galaxies are all poorly constrained. This uncertainty presents a fundamental challenge for modeling the long-term dynamical evolution of wide binaries and limits our ability to make robust predictions for their present-day populations in UFDs. More generally, using wide binaries as probes of the underlying gravitational potential inherently requires a simultaneous and self-consistent treatment of their formation and subsequent dynamical disruption. In this sense, existing predictions based on fixed primordial wide-binary populations—such as those adopted in P16—should be interpreted with caution and regarded as idealized survival prescriptions rather than complete physical models.

In light of these uncertainties, the aim of this work is not to address the formation of wide binaries or to predict their intrinsic abundance in ultra-faint dwarf galaxies. Instead, our focus is on assessing the observational capabilities of the Chinese Space Station Survey Telescope (CSST).

Looking toward the future, the CSST presents transformative opportunities to study both the large-scale structure of the universe and the resolved stellar populations of nearby galaxies. As a major flagship mission scheduled for launch in the 2027, the CSST is a two-meter aperture, off-axis, three-mirror anastigmat telescope designed for ultra-high precision cosmological surveys \citep{Zhan2011, zhan2021}. With a field-of-view (FOV) of $1.1~\mathrm{deg}^2$---approximately 300 times larger than that of the HST---and an optical quality delivering a Point Spread Function (PSF) full width at half maximum (FWHM) of less than $0.15''$, the CSST is expected to reach a limiting magnitude of $g \approx 26.3$\,mag ($5\sigma$, $2\times150$ s) \citep{nie2025,zhan2021}. During its nominal ten-year mission, the CSST will conduct a $17,500~\mathrm{deg}^2$ wide-field survey, a $400~\mathrm{deg}^2$ deep survey, and a $9~\mathrm{deg}^2$ ultra-deep-field (UDF) program, which is expected to reach a fainter limiting magnitude, e.g., $g \approx 27.5$\,mag in the $g$-band, producing seamless photometric and spectroscopic datasets for galaxies, AGNs, and supernovae \citep{Cao2018, Gong2019}. Crucially, for Milky Way satellites and nearby dwarfs within $\sim 4$\,Mpc, the CSST's high spatial resolution and survey efficiency will allow for the resolution of individual stars. Recent studies demonstrate that through effective PSF modeling, such as Multi-Gaussian fitting, the CSST can achieve astrometric centering accuracy better than 1\,mas and proper motion precision below $1.0~\mathrm{mas~yr}^{-1}$, which is realized under the condition of a multi-epoch observational strategy (5--7 epochs) over a multi-year baseline \citep{nie2025}. While this precision is insufficient to fully resolve the low internal velocity dispersion of UFDs like Segue\,I, the high-fidelity constraint is highly effective for statistically minimizing kinematic contamination from Galactic foregrounds. These capabilities offer a powerful avenue to probe binary fractions and constrain dark matter distributions in UFDs with unprecedented statistical precision \citep{Fu2023, zhan2021}.

Motivated by these challenges and opportunities, the primary aim of this work is to evaluate the capability of the CSST to detect and statistically characterize wide-binary populations in Segue\,I-like ultra-faint dwarf galaxies. Specifically, we explore how the recovery of the binary fraction depends on survey depth, sample size, and binary abundance, using the results of P16 for theoretical guidance and illustrative purposes.

In particular, we employ the two-point correlation function (2PCF) formalism to recover the observable binary fraction from projected seperation of stellar pairs under realistic observational conditions. We then examine how variations in the recovered binary fraction reflect differences in the underlying theoretical assumptions of different dark matter density profiles, thereby providing a quantitative assessment of the potential of CSST to test wide-binary–based dynamical scenarios in ultra-faint dwarf galaxies.

The structure of this paper is as follows. Section~\ref{sec:2} summarizes the physical framework governing the tidal evolution and survival of wide binaries in ultra-faint dwarf galaxies, following the $N$-body study of Peñarrubia et al.\ (2016, hereafter P16). This section is intended to provide the theoretical motivation and conceptual basis for our analysis, without introducing new dynamical modeling. In Section~\ref{sec:3}, we construct mock realizations of the ultra-faint dwarf galaxy Segue\,I by combining the wide-binary survival prescriptions derived from the P16 simulations with CSST observational constraints. We describe our binary population synthesis, mock catalog generation, and the implementation of the two-point correlation function (2PCF) formalism for recovering the binary fraction under CSST's observing conditions. Section~\ref{sec:4} presents the results of our mock observations,
quantifying CSST’s ability to recover the binary fraction in
Segue\,I analogues and to distinguish between different wide-binary
survival prescriptions. In Section~\ref{sec:5}, we discuss the advantages and limitations of wide binaries as potential tracers of the gravitational environment in ultra-faint dwarf galaxies, and outline observational strategies for a cohort of promising UFD candidates. Finally, Section~\ref{sec:6} summarizes our main conclusions.

\section{From Theoretical Disruption Models to Observable CSST Signals}
\label{sec:2}
This section details our application of the theoretical framework from P16 to forecast CSST observations. We do not perform new dynamical simulations; instead, we take the binary survival fractions for cuspy and cored dark matter halos predicted by P16's N-body simulations as our starting point. Our goal is to translate these theoretical survival probabilities into an observable signal and to assess its statistical recoverability through the two-point correlation function under realistic CSST observational conditions.
\subsection{Tidal Disruption of Binaries in UFDs}

Wide binaries, owing to their weak binding energies, are highly susceptible to tidal disruption by the host dark matter halo. This makes them effective tracers of the underlying dark matter potential \citep{Yoo2004}. In UFDs, the disruption of binary stars by the halo potential provides a key mechanism to probe the density structure of dark matter halos. These galaxies are characterized by extremely low stellar densities and high mass-to-light ratios, such that their gravitational potentials are almost entirely dominated by dark matter ($\Phi_{\rm G} \approx \Phi_{\rm DM}$). Consequently, binary stars are primarily influenced by the gradient of the dark matter potential.  

As binaries orbit within the galaxy, their components experience time-varying tidal forces from the halo. These forces act as external perturbations, injecting energy into the system. If the injected energy exceeds the binding energy ($-Gm_{\rm b}/2a$), the binary becomes unbound. The critical scale at which this occurs is quantified by the tidal radius $r_{t}$, defined as the distance within which the binary's self-gravity dominates over the external tidal field. Beyond this distance, the halo potential disrupts binary stability. The tidal radius depends sensitively on both the local dark matter density and its radial slope (i.e., $\gamma = -\mathrm{d}\log\rho/\mathrm{d}\log r$, P16), 
\begin{equation}
r_{t} = \left[ \frac{2 G m_{b} R_{0}^{2\alpha}}{5 \gamma \sigma_{0}^{2}} \right]^{1/3}
R_{h}^{2(1-\alpha)/3}
\simeq 0.27\,\mathrm{pc}
\left[ \frac{1}{\gamma(R_{h})} \cdot \frac{R_{h}}{10\,\mathrm{pc}} \right]^{1/3}.
\end{equation}
Here, \( R_h \) is the half-light radius of the dwarf spheroidal galaxy, \( R_0 = 1\,\mathrm{pc} \) , \( \sigma_0 \approx 0.93\,\mathrm{km\,s^{-1}} \) and \(\alpha \approx  0.5\) are empirical constants from the observed scaling relation \( \sigma_* = \sigma_0 (R_h/R_0)^\alpha \) \citep{walker2010}, and \( m_b = 1 M_{\odot} \) is the adopted total binary mass.
In cuspy profiles ($\gamma \sim 1$), the central density is high and declines steeply with radius, producing stronger tidal fields and hence smaller tidal radii $r_{t}$, which accelerate binary disruption even at small separations. In contrast, cored profiles ($\gamma \sim 0$) feature flat central densities and weaker tidal fields, allowing binaries to survive at larger separations. The survival fraction of binaries, and its distribution with projected separation, thus provides a direct diagnostic of whether a UFD halo follows a core or cusp profile.  

\subsection{From Binary Survival to Observable Binary Fractions}

In this study, we examine how theoretically predicted wide-binary survival models can be translated into observable quantities relevant for CSST. We consider wide binaries in different orbital configurations, characterized by their total energy (\(E\)) and angular momentum. We assume that binaries with \(E < 0\) remain gravitationally bound, which corresponds to having a semi-major axis \(a = \frac{Gm_b}{-2E} > 0\) at all times. However, it is not possible to directly measure the semi-major axis \(a\) of these binaries from observational data. Instead, we statistically infer it from the projected separation \(s\) between the stellar components on the sky.
The classical approximation $\langle s \rangle = 5\pi a / 16 \approx 0.98a$ \citep{Yoo2004} relating projected separation $s$ to the semi-major axis $a$ does not adequately describe the results in our simulations. As P16 demonstrated, for a thermal eccentricity distribution, binary orbits become radially biased during dynamical evolution, reducing the ratio $\langle s \rangle /a$ to 0.93 for cuspy halos and 0.70 for cored halos with $R_{\rm c} = 3R_{\rm h}$. However, the survival fraction curves obtained in P16 were based on the semi-major axis $a$ as the independent variable, and the relation to projected separation $s$ was only an approximation. This approximation does not fully reproduce the observed binary fractions in our simulations. 

\begin{figure}
\centering
\includegraphics[width=\columnwidth]{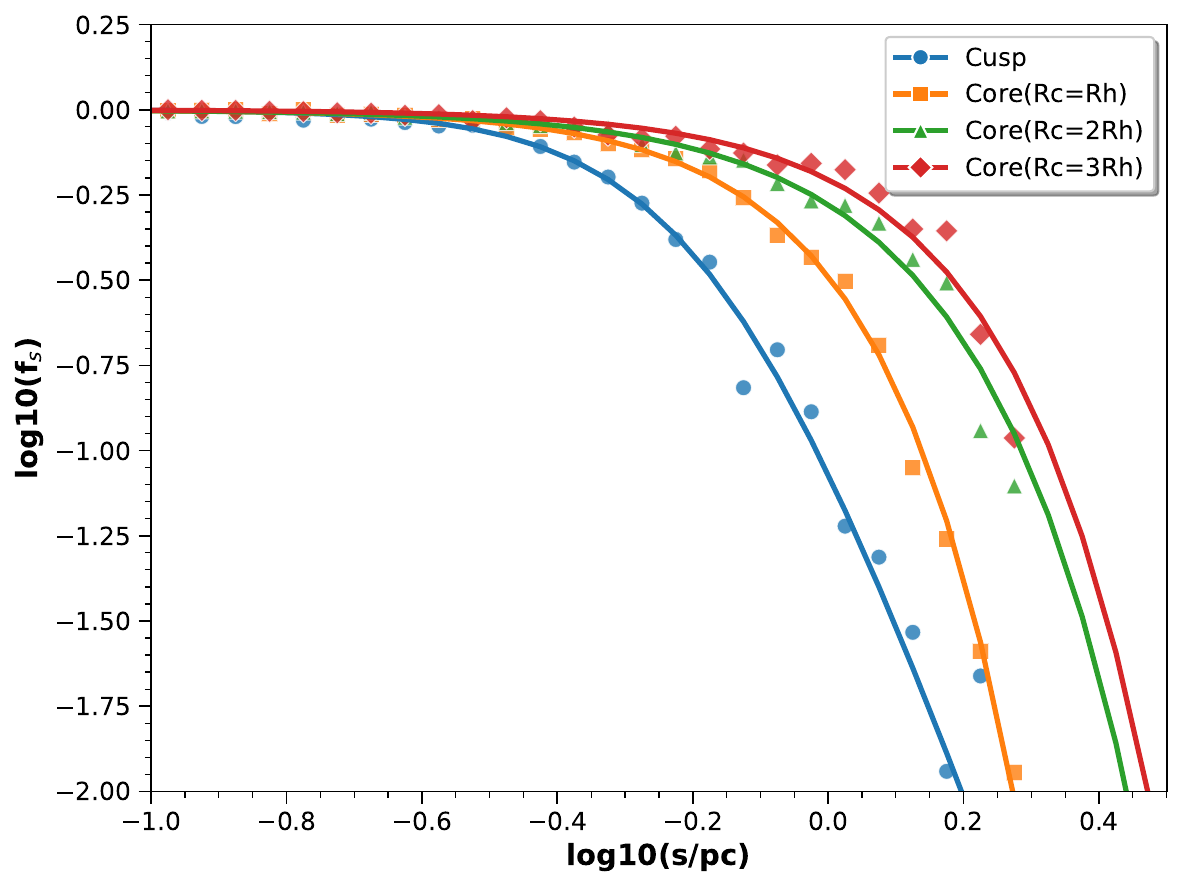}
\caption{
Binary survival fraction as a function of projected separation $s$ in Segue\,I analogues, derived from $N$-body simulations. Results are shown for cuspy (blue) and cored (red) halo profiles. The curves represent broken power-law fits to the simulation data. Compared with the analytic expectations of P16, our revised survival curves provide a more accurate recovery of the observable binary fractions in UFDs.}
\label{fig:binary_survival}
\end{figure}

To address this, we recompute the survival fraction directly as a function of projected separation, $s$, by rotating the sample multiple times to better account for the true dynamical evolution of the binaries. The updated survival curves for cuspy and cored profiles are shown in Figure~\ref{fig:binary_survival}, clearly illustrating the systematic differences in binary survival between the two halo models.

\begin{figure*}
\centering
\includegraphics[width=0.9\textwidth]{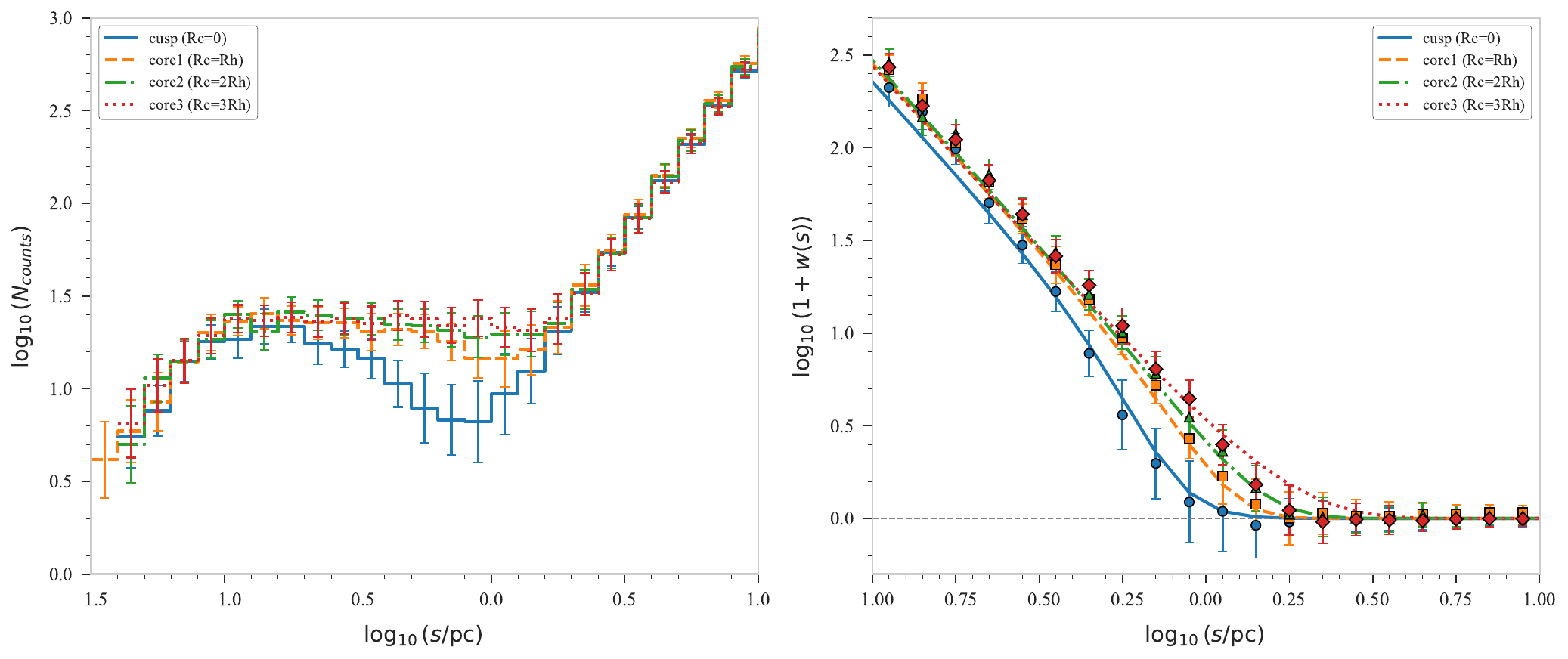}
\caption{
Comparison of stellar pair statistics in Segue\,I analogues simulated with $N_\star = 1000$ stars. \textit{Left:} Stellar pair counts as a function of projected separation. \textit{Right:} Corresponding two-point correlation functions. The cusp model ($f_{\rm b} = 0.39$) is shown in blue, while the cored models ($f_{\rm b}=0.54$--$0.63$) are shown in red. Cuspy halos exhibit a clear deficit of stellar pairs at separations $0.3$--$1$ pc, leading to systematically lower 2PCFs compared to core models. This separation range therefore provides the strongest discriminant between core and cusp dark matter profiles.}
\label{fig:correlation}
\end{figure*}

Assuming an initial binary fraction $f'_{\rm b} = 1$ at the time of galaxy formation (i.e., all stars initially in binaries), dynamical evolution produces different surviving fractions depending on the halo profile. For Segue\,I, these survival fractions correspond to the effective observable binary fraction. As results in the paper, typical survival fractions are $f_{\rm b} \approx 0.39$ for cuspy halos in Segue\,I and $f_{\rm b} = 0.54$--$0.63$ for cored halos. For example, in the cusp model, 3810 out of 10000 wide binary pairs survive, while in the core model with $R_{\rm c} = R_{\rm h}$, 5419 pairs survive; with $R_{\rm c} = 2R_{\rm h}$, 6079 pairs survive; and with $R_{\rm c} = 3R_{\rm h}$, 6361 pairs survive. These values are shown in Figure~\ref{fig:correlation}, where we compare stellar pair counts and 2PCFs for both core or cusp scenarios.

\subsection{Inferring Binary Fraction via 2PCF}

Following P16, we start by defining
the projected 2PCF of stellar pairs with separation $s$ as
\begin{equation}
1 + w(s) = \frac{\Psi(s)}{P(s)},
\end{equation}
where $\Psi(s)$ is the observed number of stellar pairs with projected separation in the interval $[s, s+ds]$, and $P(s)$ is the expectation value for a random distribution of the same $N_{\star}$ stars within the survey footprint.

To minimize projection-induced fluctuations, each simulated Segue\,I analogue is randomly rotated around three axes $N_{\rm rot}=5000$ times, and the pair separations recomputed for every projection. This procedure effectively averages over viewing angles and yields a robust estimate of the ensemble-mean $\Psi(s)$.

Wide binaries with $s \gtrsim 1$~pc are efficiently disrupted by Galactic tides and dynamical encounters, so the pair counts at $s > 1$~pc represent a purely random spatial distribution. We therefore normalize $P(s)$ to match $\Psi(s)$ at large separations, and attribute the excess signal at $s < 1$~pc entirely to surviving binaries.

For a stellar system of $N_{\star}$ stars with intrinsic binary fraction $f_{\rm b}$, the excess correlation function can be modeled as
\begin{equation}
\Psi(s) - P(s) = N_{\star} \, f_{\rm b} \, c_{s}' \, s^{-\lambda} \, f_{\rm s}(s,t),
\end{equation}
where $c_{s}'$ is a normalization constant fixed by the overall number of binary pairs, $\lambda$ is the power-law slope of the primordial separation distribution \citep[assumed $\lambda = 1$ following][]{Longhitano2010}, and $f_{\rm s}(s,t)$ is the binary survival fraction as a function of separation $s$ and system age $t$, directly measured from the $N$-body simulations. Since both $P(s)$ and $f_{\rm s}(s,t)$ are known, and $\lambda$ is assumed, the only free parameter is $f_{\rm b}$, which we determine by minimizing the $\chi^{2}$ difference between the observed excess pair counts and the model prediction.

This formalism allows a direct, model-based recovery of the intrinsic binary fraction from purely observational quantities (pair counts and survey geometry), with the dynamical survival function $f_{\rm s}(s,t)$ serving as a physically motivated correction for tidal disruption over 10~Gyr of Segue\,I's evolution.

DM substructure can introduce a significant source of degeneracy when interpreting wide binary disruption. Encounters with DM subhalos perturb and dissolve wide binaries through impulsive gravitational interactions, potentially mimicking the effects of a cuspy central density profile \citep{Yoo2004,Penarrubia2010}

Baryonic substructures may further complicate the picture. Observations of ultra-faint dwarfs such as Eridanus II reveal the presence of star clusters, suggesting that similar systems may have been present in UFDs in the past \citep{Crnjevic2016,Contenta2018}. The formation and dissolution of such clusters can inject additional dynamical heating into the stellar component, enhancing the disruption rate of wide binaries. Conversely, wide binaries can also form during the dissolution of star clusters via tidal capture or three-body interactions \citep{Peñarrubia2020}. Thus, the presence of star clusters can both disrupt and generate wide binaries, further complicating the interpretation of the observed 2PCF signal.

Disentangling these substructure effects from those arising purely from the global halo density profile requires careful modelling. We return to a more detailed discussion of these degeneracies in Section \ref{sec:5}.

\section{Simulating and Analyzing the Binary Population of Segue\,I}
\label{sec:3}

\subsection{Properties of Segue\,I}

Segue\,I is the nearest known ultra-faint dwarf galaxy. First identified in the Sloan Extension for Galactic Understanding and Exploration \citep{Belokurov2007b}, it ranks among the faintest galaxies known, with an absolute magnitude of \( M_V \approx -1.5 \) and a stellar mass on the order of \( M_\star \sim 10^3\,M_\odot \) \citep{Geha2009, Simon2011}. Spectroscopic studies reveal an extremely low mean metallicity, with [Fe/H] spanning more than 2\,dex and a characteristic value between \( -2.5 \) and \( -3.0 \) \citep{Simon2011, Frebel2014}, alongside a strong \( \alpha \)-enhancement consistent with enrichment dominated by massive stars \citep{Frebel2014}. The stellar population is essentially ancient, consistent with a brief burst of star formation early in cosmic history, with no evidence for extended or later episodes \citep{Frebel2014}.

Despite its modest stellar content, Segue\,I exhibits an exceptionally high inferred mass-to-light ratio (approximately 2000--3400 in \( M_\odot/L_\odot \)), based on dynamical modeling which indicates robust dark matter domination \citep{Geha2009, Simon2011}. Multi-epoch spectroscopy and Bayesian modeling constrain its internal velocity dispersion to \( \sigma_v \simeq 3.7^{+1.4}_{-1.1} \, \mathrm{km\,s^{-1}} \), yielding a dynamical mass within the half-light radius of \( M_{1/2} \sim 5.8 ^{+8.2}_{-3.1} \times 10^5 \, M_\odot \) \citep{Simon2011}. The half-light radius is \( r_h \approx 29^{+8}_{-5} \, \mathrm{pc} \), corresponding to a Galactocentric distance of \( \sim 23~\mathrm{kpc} \) \citep{Frebel2014}.

Segue\,I shows signs of morphological distortion, with its stellar spatial distribution appearing elongated (axis ratio \(\sim 2:1\)), leading to suggestions of potential tidal effects or extratidal features \citep{NiedersteOstholt2009, Simon2019}. However, recent comprehensive spectroscopic surveys find no conclusive kinematic evidence for ongoing tidal disruption and have reduced earlier estimates of contamination from the Sagittarius stream, affirming its status as a dark matter-dominated dwarf galaxy \citep{Simon2011}. Given its extreme dark matter density, compact size, and resolved stellar dynamics, Segue\,I offers a unique and stringent environment for probing the relationship between stellar tracers, dark matter distribution, and environmental perturbations. The survival (or disruption) of wide binaries in such an environment is particularly sensitive to these underlying gravitational potentials and any tidal heating (P16), making Segue\,I a prime system for constraining these physical processes.

In the context of Segue\,I, we adopt the binary evolution framework from P16, using $N$-body simulations to model the dynamical evolution of binary stars within the galaxy's dark matter halo. This modeling approach accounts for eccentric binary orbits, anisotropic velocity distributions, and the evolving nature of the dark matter halo. The core properties of Segue\,I, including its mass-to-light ratio, velocity dispersion, and spatial distribution, are key to determining how wide binaries may survive or be disrupted by the galaxy's internal dynamics.

\subsection{Binary Population Synthesis in Mock Segue\,I}
\label{sec:3.2} 
In order to investigate the impact of binary stars on the observed properties of Segue\,I, we constructed a mock stellar catalogue based on the $N$-body simulations of P16, which reproduce the dynamical structure and kinematics of this ultra-faint dwarf galaxy with high fidelity. From the simulation data, we randomly selected a subsample of $N_{\star}=1000$ tracer particles to serve as the basis of our synthetic stellar population. Each selected particle was then classified either as a single star or as a binary system according to a prescribed binary fraction,
\begin{equation}
f_{\rm b} = \frac{N_{\rm binary}}{N_{\star}},
\end{equation}
where $N_{\rm binary}$ is the number of binary stars and $N_{\star}$ is the total number of stars in the sample.  The orbital properties of wide binaries in our mock catalog are inherited from P16. In those simulations, the initial binary population is characterized by a power-law semimajor-axis distribution $q(a,t=0)\propto a^{-\lambda}$, with $\lambda=1$, spanning the range $a_{\rm min}=0.02$~pc to $a_{\rm max}=2.0$~pc. For a total binary mass of $1\,M_\odot$. The eccentricity distribution is assumed to be either random or thermal, reflecting current observational uncertainties.

To estimate the total number of stars in Segue\,I and construct its stellar mass function, we first perform a benchmark calculation using the classical Salpeter initial mass function \citep[IMF,][]{Salpeter1955}. The Salpeter IMF is expressed as a single power law within the applicable mass range, with a slope index of $\alpha = -2.35$. Assuming a lower stellar mass limit of $0.08~M_{\odot}$ and constraining the total stellar mass of Segue\,I to approximately $\sim 1300\,M_{\odot}$ \citep{Simon2011}, we estimate that the galaxy contains roughly 3000 stars. This estimate provides a first-order approximation for the stellar population of Segue\,I.

However, numerous studies suggest that the IMF may deviate from a single Salpeter slope at the low-mass end. To refine our model, we adopt a combined approach using the multi-segment \citep{Kroupa2001} IMF together with empirically constrained IMF measurements in UFDs. The Kroupa IMF provides a piecewise power-law description with slopes of
\[
\alpha_1 = 1.3 \quad (0.08 < m/M_\odot < 0.5), \quad 
\alpha_2 = 2.3 \quad (0.5 < m/M_\odot),
\]
which effectively reproduces the stellar mass distribution in the Milky Way disk.  

Deep Hubble Space Telescope/ACS observations of several UFDs reveal significant deviations from this Galactic IMF. \citet{Gennaro2018} analysed six UFDs (including Bo\"otes\,I, Leo\,IV, and Hercules) and found systematically bottom-light IMFs over the mass range $0.45 < m/M_\odot < 0.8$, with single-slope fits ranging from $\alpha \approx 1.0$ (Leo~IV) to $\alpha \approx 1.9$ (Bo\"otes\,I), much flatter than the Salpeter value ($\alpha = 2.35$). Similarly, \citet{Geha2013} measured $\alpha = 1.2^{+0.4}_{-0.5}$ for Hercules and $\alpha = 1.3 \pm 0.8$ for Leo\,IV in the mass interval $0.52 < m/M_\odot < 0.77$. These results indicate a relative deficiency of low-mass stars compared to the Milky Way field population, implying that the IMF in UFDs is environmentally dependent, being shallower (more bottom-light) in low-metallicity, low–velocity dispersion systems.

Moreover, recent theoretical developments based on the integrated galaxy-wide IMF (IGIMF) framework \citep{Yan2024} predict that UFDs, owing to their extremely low star-formation rates and metallicities, should exhibit both top-light and bottom-light IMFs. This agrees with the empirical trends observed across Local Group dwarfs, where IMF slopes become progressively shallower with decreasing metallicity and galactic potential depth. Taken together, these results suggest that the IMF of Segue\,I likely deviates from universality, favoring a bottom-light form consistent with the observed properties of other UFDs systems. In our study, we adopt $\alpha = 1.2$. 

For single stars, we directly inherit the phase-space coordinates of the parent tracer particle, and assign stellar masses drawn from the IMF. For binary stars, the primary star mass $m_1$ is sampled from the same IMF, while the companion mass $m_2$ is determined through the mass ratio $q = m_2/m_1$, such that the total system mass satisfies $m_1+m_2=1~M_{\odot}$. 

To explore the sensitivity of our results to the binary mass-ratio distribution, we adopt four functional forms widely used in the literature: (i) a normal distribution, (ii) an exponential distribution, and (iii) a negative exponential distribution. The probability density functions of these models follow those presented by \citet{Li2024}, whose Figure\,5 explicitly shows the adopted forms:
\begin{align}
    f_{\rm normal}(q) &= \frac{1}{\sqrt{0.08\pi}} \exp\!\left[-\frac{(q-0.5)^{2}}{0.08}\right], \\
    f_{\rm exp}(q) &= 3 e^{-3q}, \\
    f_{\rm negexp}(q) &= 1 - 3 e^{-3q},
\end{align}
where $q \in [0,1]$. In addition to these analytic prescriptions, we also adopt an observationally-motivated mass-ratio distribution based on solar-neighbourhood binaries, as measured by \citet{ElBadry2019}. This model is described by a broken power-law distribution with an excess population of ``twin'' binaries near $q \simeq 1$:
\begin{equation}
p(q) \propto
\begin{cases}
q^{\gamma_{\rm smallq}}, & 0.3 < q < q_{\rm break},\\[6pt]
q^{\gamma_{\rm largeq}}, & q_{\rm break} \le q < q_{\rm twin},\\[6pt]
(1+F_{\rm twin}) q^{\gamma_{\rm largeq}}, & q_{\rm twin} \le q \le 1,
\end{cases}
\end{equation}
where $\gamma_{\rm smallq}$ and $\gamma_{\rm largeq}$ are the logarithmic slopes below and above $q_{\rm break}$, respectively. $F_{\rm twin}$ is the excess fraction of nearly equal-mass binaries with $q>q_{\rm twin}$ relative to the underlying power-law distribution, and typical parameter values are $\gamma_{\rm smallq}=0.3$, $\gamma_{\rm largeq}=-1.3$, $q_{\rm break}=0.5$, $q_{\rm twin}=0.95$, and $F_{\rm twin}=0.04$.  

\subsection{Mock Photometric properties in CSST}

\begin{figure}
  \centering
  \includegraphics[width=\linewidth]{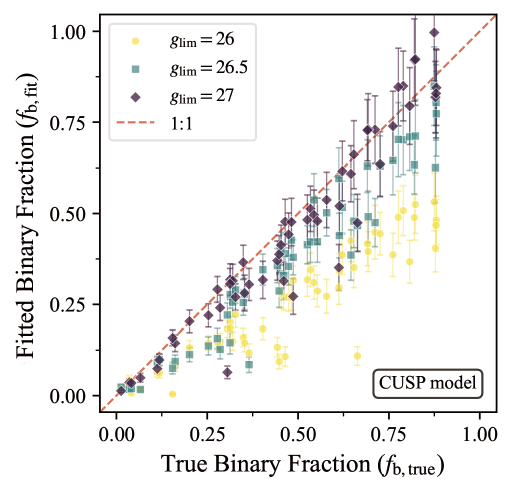}
  \caption{
  Recovery of the binary fraction ($f_\mathrm{b}$) for the cusp model ($R_\mathrm{c} = R_\mathrm{h}$) with varying $g$-band limiting magnitudes. The fitted binary fraction ($f_{\mathrm{b,fit}}$) is plotted as a function of the true binary fraction ($f_{\mathrm{b,true}}$). Data points are color-coded by the limiting $g$-band magnitude, with yellow representing $g_\mathrm{lim} = 26$, green for $g_\mathrm{lim} = 26.5$, and purple for $g_\mathrm{lim} = 27$. The error bars denote the 95\% confidence intervals for each measurement. The dashed red line represents the one-to-one relation. At the deepest field depth ($g \sim 27$ mag), the recovery is nearly unbiased, while shallower limits ($g \sim 26$ mag) introduce a slight underestimation of $f_\mathrm{b}$ due to incompleteness effects.}

  \label{fig:fb_recovery}
\end{figure}

Photometric properties are then assigned by interpolating PARSEC isochrones \citep{Bressan2012,Chen2014,Marigo2017}, assuming an age of $10~{\rm Gyr}$, which is consistent with previous studies of UFDs \citep{Brown2014}, and a metallicity of $\mathrm{[Fe/H]}=-2.0$. For each star, its $g$-band magnitude is obtained by linearly interpolating along the isochrone according to its mass. This procedure yields a synthetic stellar catalogue that is dynamically consistent with Segue\,I. The fidelity of this binary population synthesis and recovery method is quantitatively assessed in Figure~\ref{fig:fb_recovery}, which illustrates the recovery of the intrinsic binary fraction \( f_{\rm b} \) for our simulated Segue\,I analogues. This analysis explores the interplay between the input binary fraction, the adopted \(g\)-band limiting magnitude, and the assumed mass-ratio distribution (broken power-law with twin excess, uniform, exponential, and negative exponential). The results demonstrate that at the deep photometric limit expected for CSST deep fields (\(g \sim 27.5\)\,mag), the recovery of \( f_{\mathrm{b}} \) is nearly unbiased across the full range of input fractions and for all mass-ratio models tested. In contrast, under shallower survey conditions (\(g \sim 26.0\)\,mag), the incompleteness in detecting faint binaries introduces a systematic bias, leading to a slight underestimation of the intrinsic binary fraction. The shaded bands, representing the scatter across multiple realizations, further quantify the precision of our method, confirming that the binary fraction in low-density systems can be constrained with deep, high-precision astrometry.

\subsection{Stellar Population in CSST Mock Catalog}
To generate a realistic mock sample of Segue\,I stars for our analysis, we make use of the first comprehensive Milky Way stellar mock catalog for the CSST Survey Camera presented by \citep{Chen2023}. This catalog, constructed using the \textsc{TRILEGAL} population synthesis tool, contains $\sim 12.6$ billion stars with self-consistent photometry, astrometry, and kinematics, reaching down to $g = 27.5$~mag in the AB system. From this catalog, we spatially select stars within the sky region corresponding to the known extent of Segue\,I. The resulting stellar sample serves as the single-star component of our mock Segue\,I analogs.

We then combine this mock stellar field with our binary population synthesis described in Section \ref{sec:3.2}. Specifically, we randomly select $N_{\star}=1000$ stars from the Segue\,I region of the CSST mock catalog and stochastically assign them as single stars or binary stars according to a prescribed input binary fraction $f_{\rm b}$. The binary components are drawn from the mass distributions contained twin binaries described in Section \ref{sec:3.2}. This procedure yields a self-consistent Segue\,I stellar population model including both single and binary stars.

\begin{figure}
  \centering
  \includegraphics[width=\linewidth]{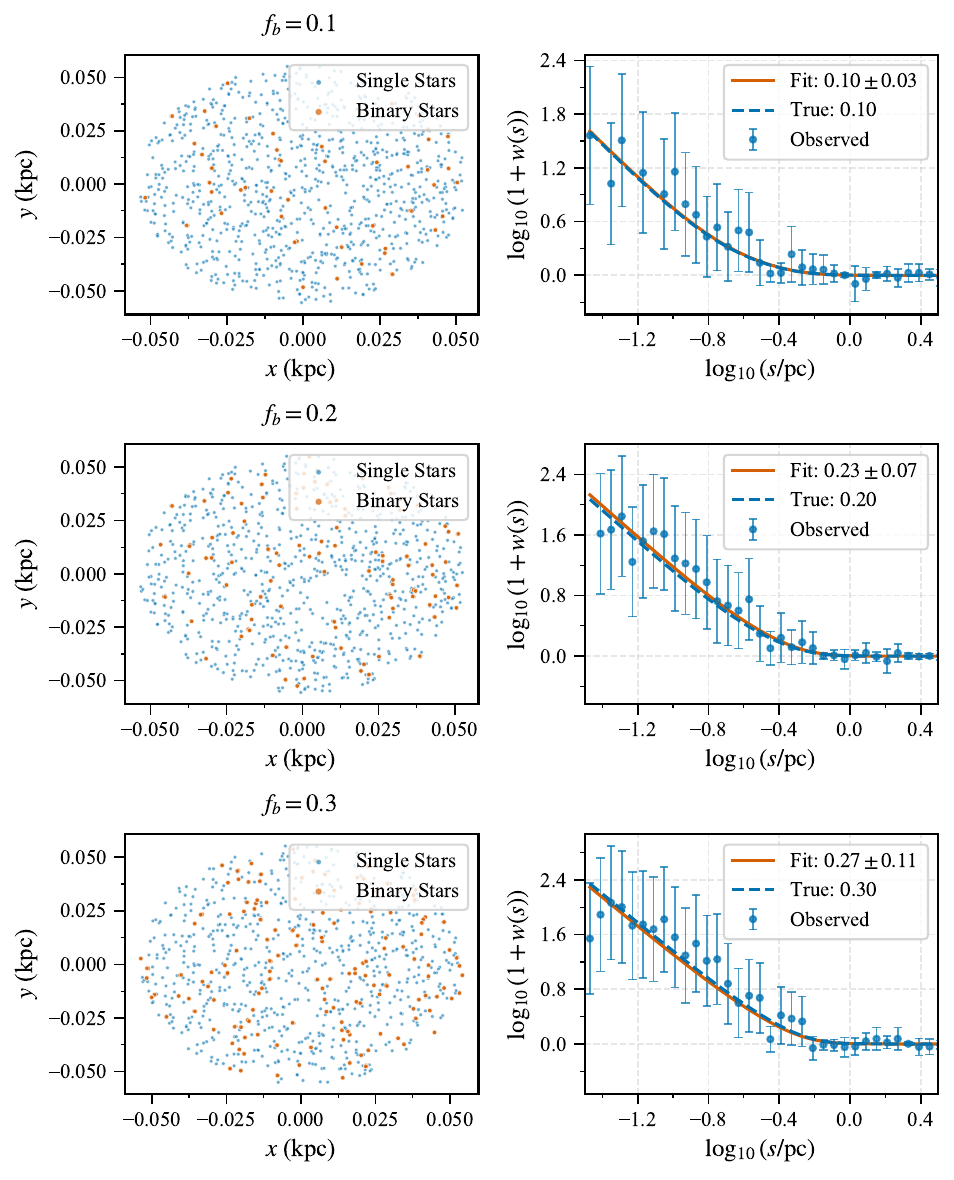}
  \caption{Spatial distributions (left panels) and two-point correlation functions (right panels) for simulated stellar systems with varying binary fractions $f_b$. Left: projected stellar positions, color-coded for single (blue) and binary (red) populations. Right: Observed $\log_{10}(1+w(s))$ versus $\log_{10}(s/\mathrm{pc})$, compared to best-fit models and true input values (dashed lines). Increasing $f_b$ enhances small-scale clustering, while recovered $f_b$ values remain consistent with input parameters within uncertainties.}
  \label{fig:csst_recovery}
\end{figure}

%This combined modeling framework is exemplified in Figure~\ref{fig:csst_recovery}.{\magenta JP: typo} The efficacy of our method in constraining the binary fraction is quantitatively demonstrated in Figure~\ref{fig:csst_recovery}. 
This combined modeling framework is exemplified in Figure~\ref{fig:csst_recovery}, which also quantitatively demonstrates the efficacy of our method in constraining the binary fraction. This figure presents a comparative analysis of mock stellar populations generated with known input binary fractions ($f_b = 0.1, 0.2, 0.3$). The left panels display the projected spatial distributions of stars, where single stars and binary systems are distinctly color-coded in blue and red, respectively. A visual inspection of these spatial maps reveals a perceptible increase in small-scale clustering with higher binary fractions. The right panels provide a quantitative measure of this clustering through the two-point correlation function, plotted as $\log_{10}(1+w(s))$ against $\log_{10}(s/\mathrm{pc})$. The observed correlation functions (data points) show a pronounced enhancement at small separations for systems with larger $f_b$. Crucially, the best-fit models (solid lines) excellently reproduce the observed data and consistently yield binary fraction estimates that are in remarkable agreement with the true input values (indicated by the dashed vertical lines), all within the statistical uncertainties. This result robustly validates the reliability of our analysis pipeline in accurately recovering the binary fraction from astrometric data.

\subsection{Observational Limits}

The sensitivity of the 2PCF to the underlying binary population is strongly influenced by the observational capabilities of the survey. Two factors play a dominant role: angular resolution and photometric depth. The angular resolution sets the minimum resolvable projected separation that contributes to the 2PCF signal. For instance, at the distance of Segue\,I, the $0.15''$ spatial resolution of CSST corresponds to a projected physical separation of $\sim 0.02$~pc, below which binary components become unresolved. Consequently, our mock 2PCF analysis is insensitive to binaries with separations smaller than this threshold, but unaffected for wider systems.  

\begin{figure}
  \centering
  
  \includegraphics[width=0.95\linewidth]{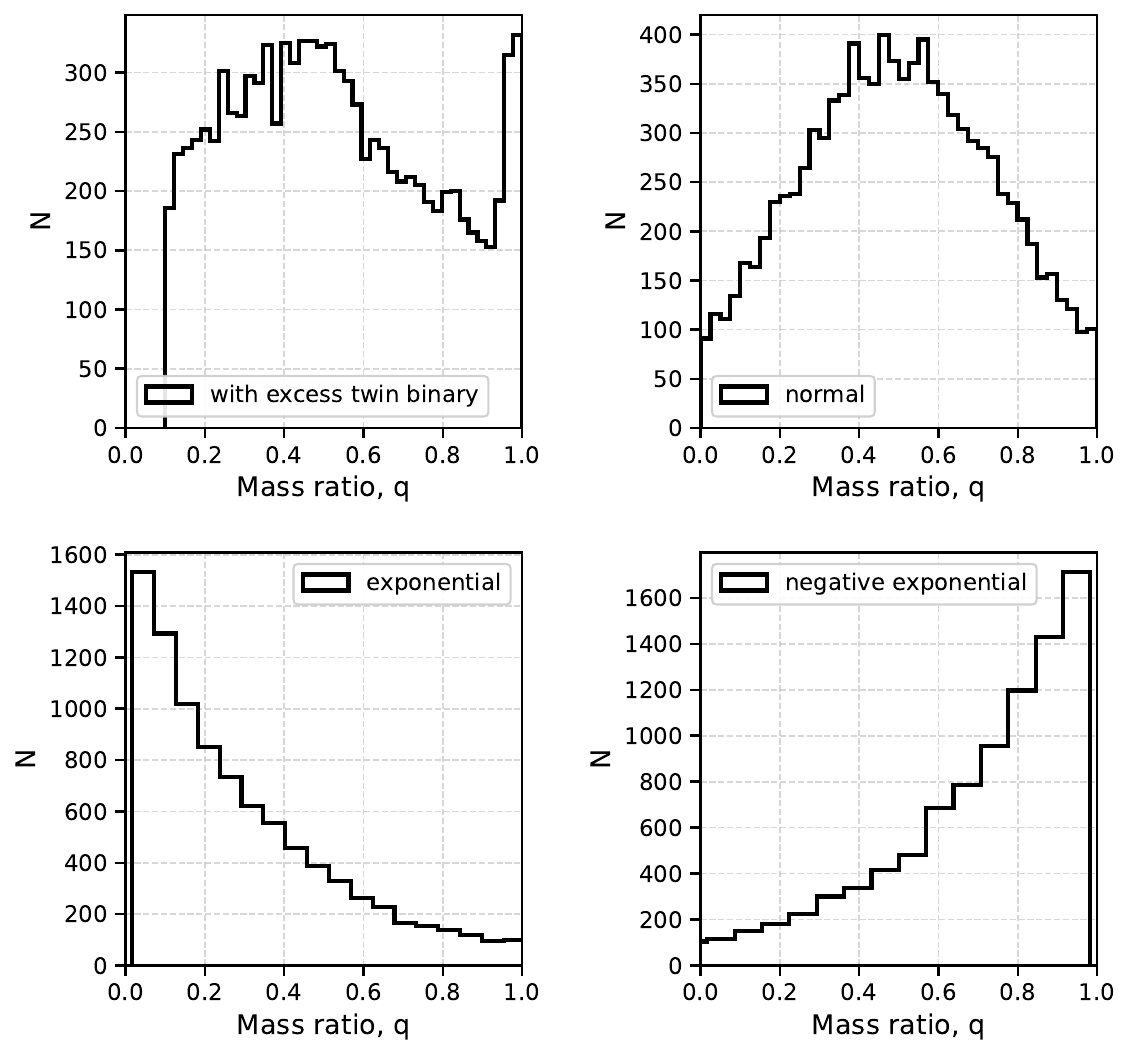}
  \caption{Binary star mass-ratio distributions considered in this study, including a broken power-law with ''twin'' binary excess \citep{ElBadry2019}, a uniform distribution, and exponential/negative-exponential forms \citep{Li2024}.}
  \label{fig:mass_ratio_distributions}
  
  \vspace{0.2cm} % 添加间距，可根据需要调整
  
  \includegraphics[width=0.95\linewidth]{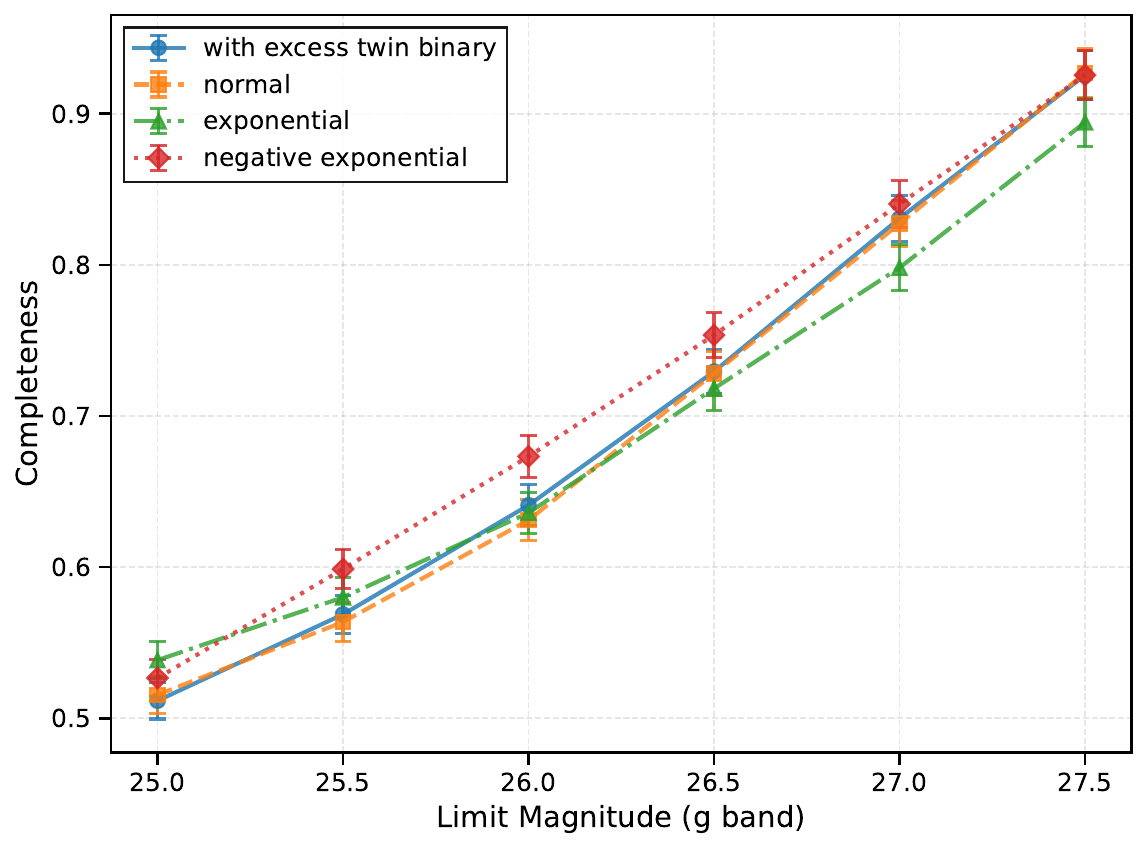}
  \caption{Detection completeness as a function of $g$-band limiting magnitude for the mass-ratio distributions shown above. Completeness rises steeply with increasing depth, reaching $\sim 90\%$ at $m_{g} \approx 27.5$. Equal-mass binaries are preferentially detected, resulting in a $\sim 15\%$ higher completeness relative to distributions dominated by low-$q$ systems at the faintest limits.}
  \label{fig:completeness_plot}
\end{figure}

\begin{figure}
    \centering
    \includegraphics[width=0.8\columnwidth]{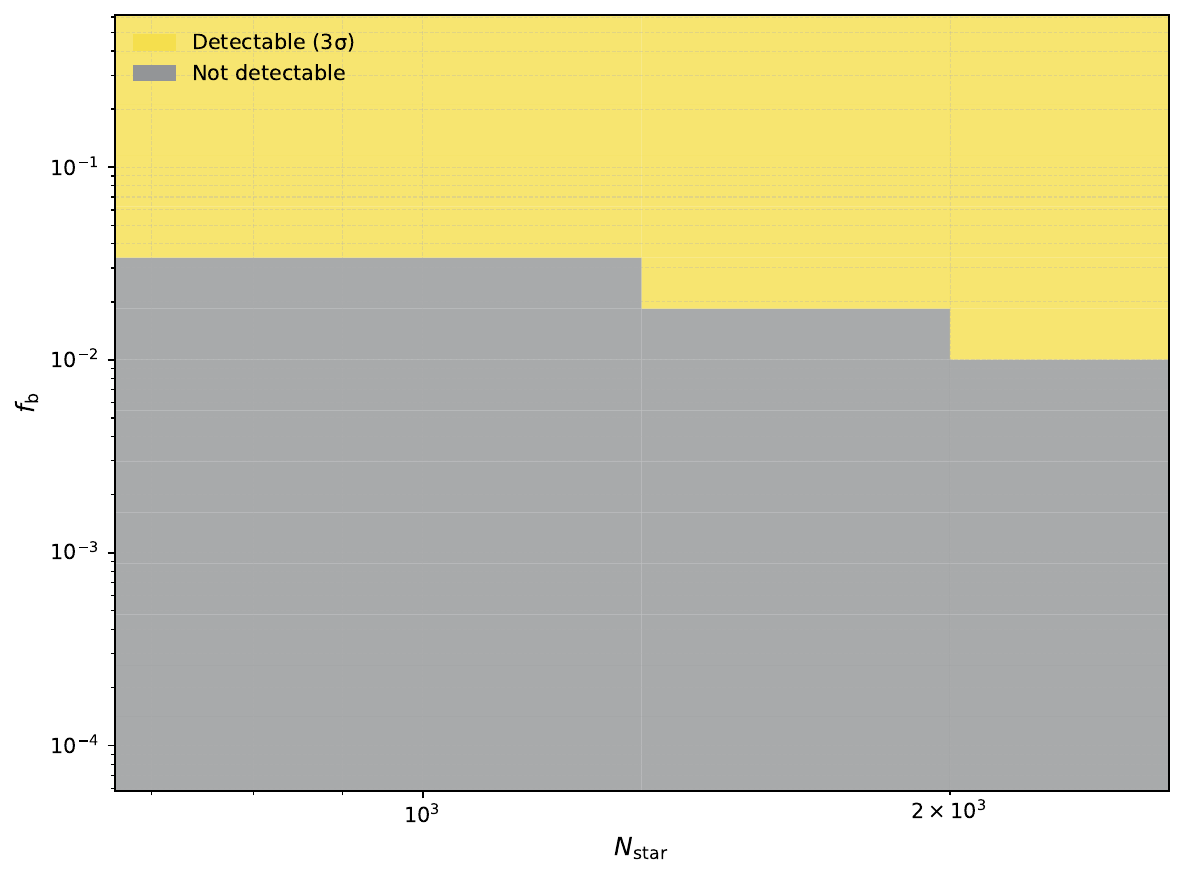}
    \caption{Detectability map for wide binaries in the $N_{\mathrm{star}}$--$f_{\mathrm{b}}$ plane at a distance of Segue\,I. Yellow regions indicate parameter combinations where the $3\sigma$ detection criterion is satisfied, while grey regions denote non-detections. The results show that for $N_{\mathrm{star}} \approx 1000$, a binary fraction $f_{\mathrm{b}} \gtrsim 0.04$ is required for a detectable signal. This threshold decreases to $f_{\mathrm{b}} \gtrsim 0.024$ for $N_{\mathrm{star}} \approx 1700$ and further to $f_{\mathrm{b}} \gtrsim 0.013$ for $N_{\mathrm{star}} \approx 2300$.}
    \label{fig:detectability}
\end{figure}
Thanks to its wide $1.1^{\circ} \times 1.1^{\circ}$ field of view, CSST can cover the entire spatial extent of typical UFDs (angular sizes $\lesssim 0.5^{\circ}$) in a single pointing, enabling simultaneous measurement of both their dense stellar cores and diffuse outer halos. This is a significant advantage compared to the \textit{Hubble Space Telescope} (HST), whose Advanced Camera for Surveys (ACS) and Wide Field Camera 3 (WFC3) provide a field of view of only $\sim 3' \times 3'$, requiring mosaics of multiple pointings to cover similar angular areas. The large areal coverage of CSST not only improves observational efficiency but also allows for robust measurement of the random pair baseline, which is critical for an accurate determination of the 2PCF.

The limiting magnitude determines the level of photometric completeness. Deeper observations increase the number of detected stars and random pairs, thereby lowering the baseline $P(s)$ and enhancing the amplitude of the 2PCF on small scales. To quantify the impact of binary mass-ratio distributions on detectability, we compute the completeness fraction as a function of $g$-band limiting magnitude for four different $q$-distributions. Figure~\ref{fig:mass_ratio_distributions} shows the adopted mass-ratio distributions, while Figure~\ref{fig:completeness_plot} displays the corresponding completeness curves. We find that the completeness fraction is only weakly sensitive to the assumed $q$-distribution: for a limiting magnitude of $m_{g} \approx 26$, the overall completeness reaches $\sim 63\%$, increasing to $\sim 90\%$ at $m_{g} \approx 27.5$. Systems with mass ratios near unity are $\sim 15\%$ more likely to be detected than those with strongly suppressed low-$q$ companions at $m_{g} = 27.0$, consistent with the enhanced brightness of equal-mass binaries.

Our simulations investigated the observational feasibility of distinguishing the Dark Matter core and cusp profiles in UFDs by utilizing the dynamical signatures of binary stars, under the constraints of the CSST limiting magnitude ($g$-band $\sim 27.5$ mag). The core finding is a strong interdependence among the critical parameters for observational success: the UFD distance ($D$), the total number of observed stars ($N_{\text{star}}$), and the initial binary fraction ($f'_b$). 

\begin{figure*} % 使用 figure* 环境确保图表跨越两栏（占据一页宽度）
    \centering

    % --- 子图 1: N_star = 1000 ---
    \centering
    % 使用 \linewidth 确保图片占据 subfigure 的全部宽度
    \includegraphics[width=\linewidth]{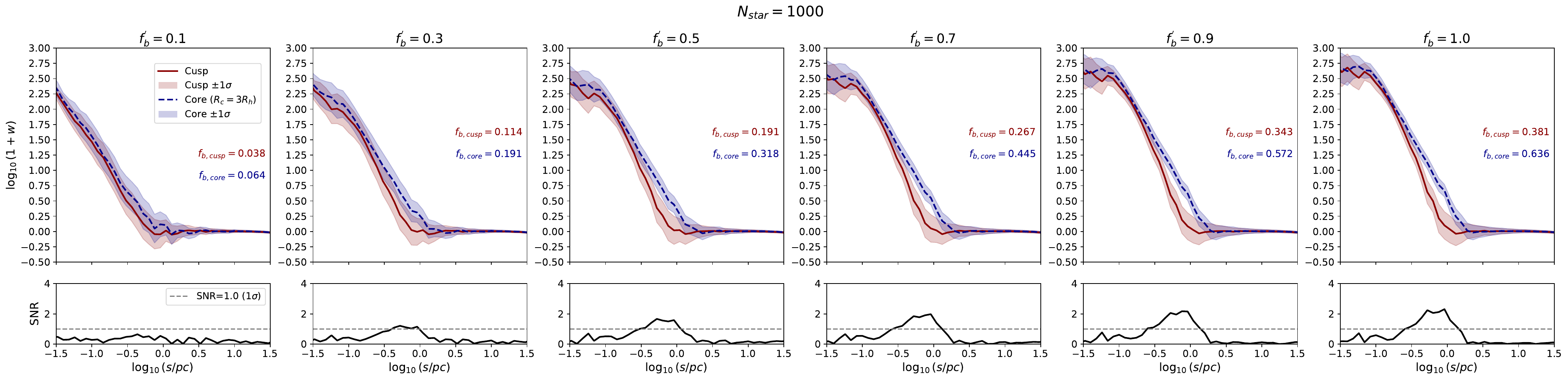} 
        
    \label{fig:n1000}
    
    \vspace{0.1cm} % 增加垂直间距，保持视觉分离
    
    % --- 子图 2: N_star = 2000 ---
    \centering
    \includegraphics[width=\linewidth]{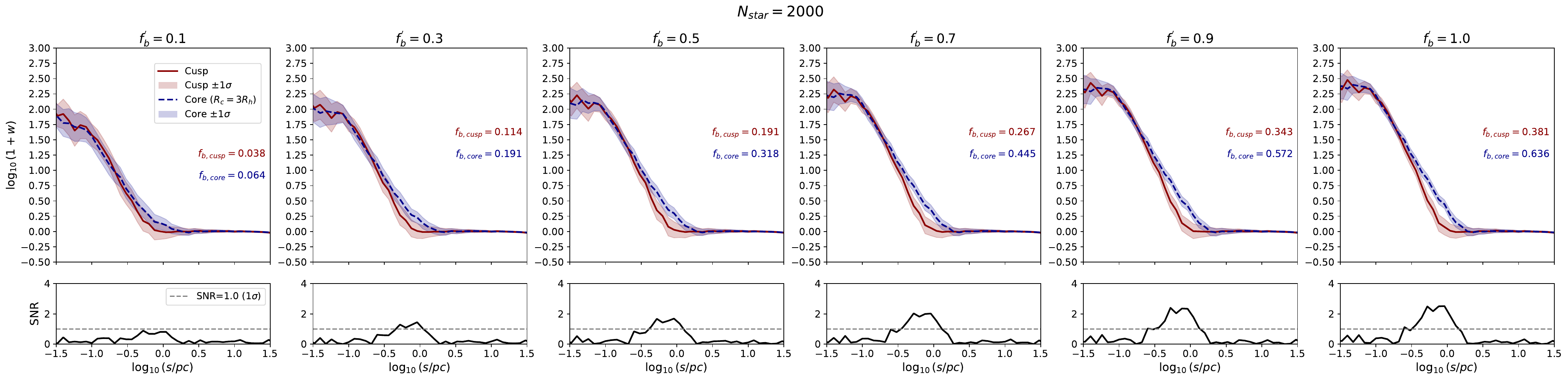}
        
    \label{fig:n2000}
    
    \vspace{0.1cm} 
    
    % --- 子图 3: N_star = 3000 ---
    \centering
    \includegraphics[width=\linewidth]{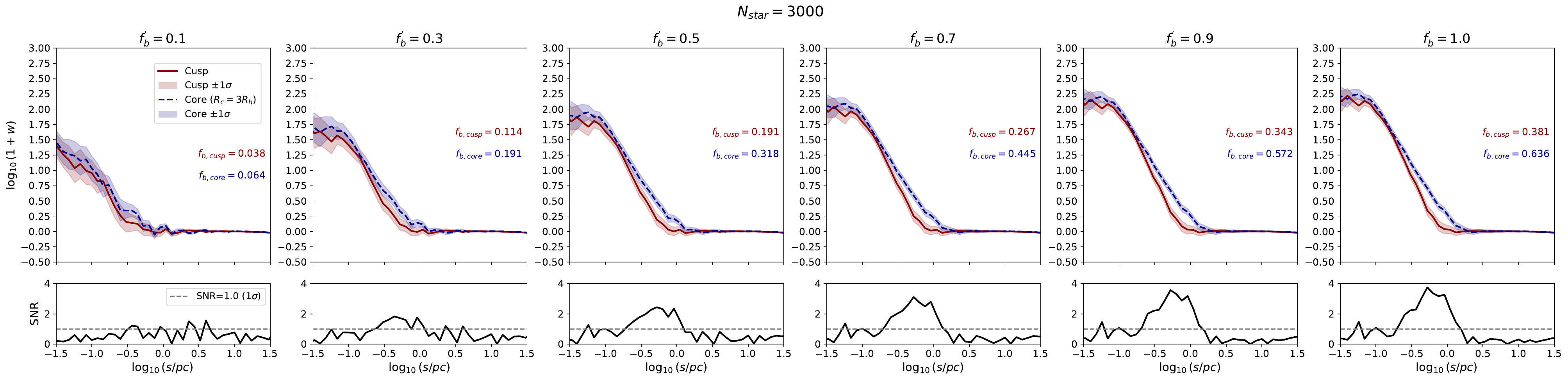}
    
    \label{fig:n3000}
    
    % --- 总图注 (必须在图表下方) ---
    \caption{Projected two-point correlation functions and their relative separation for cusp and core halo models, as derived from the P16 $N$-body simulations.
    The figure explores a range of hypothetical stellar sample sizes ($N_{\rm total}$) and assumed initial binary fractions ($f'_b$) in order to map the parameter space in which differences between wide-binary survival prescriptions would become statistically separable in principle.
    Panels are arranged vertically by total stellar population: $N_{\rm total}=1000$ (top row), $2000$ (middle row), and $3000$ (bottom row). Columns correspond to different assumed values of $f'b$. The fractions of binaries surviving dynamical disruption are given by $f{b,\mathrm{cusp}}$ and $f_{b,\mathrm{core}}$ for the cusp and core models, respectively.
    Top panels: The projected two-point correlation function $\log_{10}(1+w)$ for the cusp (solid red) and core ($R_c=3R_h$, dashed blue) models. Shaded regions indicate the $\pm1\sigma$ scatter across realizations.
    Bottom panels: A model-based separability metric defined as ${\rm SNR} = |\log_{10}(1+w){\rm cusp} - \log{10}(1+w){\rm core}| / \sigma{\rm joint}$, shown as a function of projected separation $s$. This quantity is not an observational statistic, but is used here to quantify the relative divergence between survival prescriptions under controlled assumptions. The dashed grey line marks ${\rm SNR}=1$ as a reference level.
    }
    \label{fig:all_samples}
\end{figure*}

\section{Results}
\label{sec:4}

A central finding of this study is the quantitative confirmation that the 2PCF possesses sufficient statistical power to recover the binary fraction in UFDs. 
We have carried out extensive Monte Carlo simulations to assess the detectability of wide binaries as a function of the stellar sample size $N_{\text{star}}$ and the intrinsic binary fraction $f_b$. For each combination of these parameters, we generated 50 independent realisations of a Segue\,I--like stellar population, accounting for the distance modulus, CSST's limiting magnitude, and the resolution limit.

The projected 2PCF was computed for each mock catalogue, and the binary fraction was recovered by fitting the small-scale excess with the model described in Section~\ref{sec:3.2}. A detection was deemed successful when the fitted value $\hat{f}_b$ satisfied the $3\sigma$ criterion: the bias (mean of $\hat{f}_b$ over realisations minus the true $f_b$) is smaller than three times the standard deviation of $\hat{f}_b$, and the standard deviation itself is less than 50\% of $f_b$ to ensure a meaningful estimate.

Figure~\ref{fig:detectability} shows the resulting detectability maps in the $N_{\text{star}}$--$f_b$ plane for both distances. The colour scale indicates whether the $3\sigma$ condition is met (yellow) or not (grey). For Segue\,I at 23 kpc, a sample of $\approx 1000$ stars requires $f_b \gtrsim 0.04$ for a detectable wide-binary signal. This threshold decreases to $f_b \gtrsim 0.024$ for $N_{\text{star}} \approx 1700$ and further to $f_b \gtrsim 0.013$ for $N_{\text{star}} \approx 2300$. At the larger distance of 30 kpc, the increased distance modulus reduces the number of observable stars for a given input $N_{\text{star}}$, shifting the detection thresholds to higher binary fractions. For example, with $N_{\text{star}} \approx 2000$, binaries become detectable only when $f_b \gtrsim 0.1$, while for $N_{\text{star}} \approx 3500$ the threshold decreases to $f_b \gtrsim 0.05$. These values are consistent with the simple expectation that a larger stellar population reduces Poisson fluctuations and sample variance, thereby allowing fainter binary signals to be extracted, while greater distance dims the observable sources and raises the required binary fraction.

The above results demonstrate that, even without a detailed understanding of wide-binary formation, CSST will be capable of probing binary fractions at the percent level in ultra-faint dwarfs similar to Segue\,I, provided several thousand stars are observable. For more distant UFDs ($\gtrsim 30$ kpc), the required binary fractions are higher ($\sim 5-10$ per cent), but still within theoretically plausible ranges.

While the results above establish the feasibility of identifying a wide-binary population with CSST, the ultimate scientific goal is to use these binaries as dynamical probes of the underlying dark-matter potential.We quantify this discriminative power using the cumulative signal-to-noise ratio (SNR), calculated from the difference in the 2PCF amplitudes between cusp and core models across the sensitive separation range of $0.3 - 1$\,pc.

Our simulations reveal a clear and positive dependence of the SNR on both the assumed initial binary fraction ($f_b'$) and the stellar sample size ($N{\rm star}$). This behaviour is illustrated in Fig.~\ref{fig:all_samples}, which summarizes how the cumulative SNR varies across the explored parameter space. For a fixed stellar sample size, the SNR increases monotonically with $f'_b$, reflecting the fact that a larger binary fraction enhances the excess of close stellar pairs and therefore strengthens the small-scale clustering signal measured by the 2PCF.

For a fixed initial binary fraction, the SNR exhibits a strong scaling with $N_{\rm star}$. For example, at $f_b' = 0.1$, the SNR increases from 3.18 for $N{\rm star}=1000$ to 5.94 for $N_{\rm star}=3000$. At higher binary fractions, a similar trend is observed: for $f_b' = 0.5$, the SNR grows from 4.61 to 7.53, and for $f_b' = 0.9$, from 5.62 to 9.15 over the same range in $N{\rm star}$. This behaviour is broadly consistent with the expected $\sqrt{N{\rm star}}$ scaling of pair-count statistics, indicating that the statistical sensitivity of the method is driven primarily by sample size.

We emphasize that the SNR shown in Fig.~\ref{fig:all_samples} does not represent an observational statistic that can be directly measured from a single galaxy. Instead, it should be interpreted as a model-based separability metric, quantifying the degree to which the cusp and core survival prescriptions derived from the P16 $N$-body simulations diverge under controlled assumptions. The figure is therefore intended to illustrate the statistical detectability in principle of differences between survival models, rather than to provide a definitive criterion for distinguishing dark-matter density profiles in an individual system.

We further note that a substantial fraction of the parameter space shown in Fig.~\ref{fig:all_samples} is unlikely to be directly realized in Segue\,I itself. In particular, the observable wide-binary fraction is expected to be significantly lower than the adopted initial binary fraction $f'_b$, and the number of stars accessible to CSST in Segue\,I is unlikely to reach the highest values explored here. The motivation for including this extended parameter space is therefore not to make a realistic prediction for Segue\,I, but rather to identify approximate thresholds in stellar sample size and binary abundance beyond which differences between theoretical survival prescriptions would, in principle, become detectable.

To establish general feasibility criteria, we simulated the 2PCF signal of Segue\,I at various assumed distances. The results indicate that within $D \lesssim 40$~kpc, distinguishing between cusped and cored dark matter profiles at the $1\sigma$ confidence level becomes feasible when the stellar sample size reaches $N_{\text{star}} \gtrsim 6000$ and the binary fraction exceeds $f_{\text{bin}} \gtrsim 0.1$. It is important to note that while this study adopts a mean SNR-based estimate, a more realistic cumulative SNR under deep, multi-band imaging conditions will substantially improve sensitivity, implying that the practical feasibility for profile discrimination is likely even higher. Furthermore, the SNR scales approximately with $\sqrt{N_{\text{star}}}$, indicating that the method's statistical power will grow substantially with larger stellar samples from future deep surveys.

\section{Discussion}
\label{sec:5}
\subsection{Limitations and Future Refinements}

While our results are highly promising, we identify several avenues for future refinement that will enhance the realism and robustness of our method when applied to actual observational data.

First, our present model operates on the N-body simulation. In practice, contamination from foreground Milky Way stars and background galaxies is an inevitable challenge. Although the CSST's excellent spatial resolution and deep photometry will aid in member selection through color-magnitude diagram filtering, the residual contamination could potentially bias the measured 2PCF. A natural extension of this work is to incorporate a realistic contamination model into the simulations and to develop robust statistical background subtraction techniques for the 2PCF measurement.

Second, our analysis has not yet propagated the impact of astrometric uncertainties. The finite precision in measuring stellar positions, especially for faint stars at the limiting magnitude of the survey, will act to smear out the closest pairs, potentially suppressing the 2PCF signal at the smallest separations most critical for our analysis. A thorough investigation, folding a detailed CSST astrometric error model into our mock observations, is required to quantify this effect and to determine the true smallest resolvable separation for the wide binary population.

Finally, the current model assumes a pure wide binary population. The presence of hierarchical triple or quadruple systems, as well as chance alignments of unrelated stars, could introduce a non-dynamical signal. Future work will involve testing the sensitivity of our method to these astrophysical confounders, ensuring that the recovered binary fraction and the subsequent dark matter constraints are not significantly biased.
\begin{table*}
\centering
\caption{Basic properties of selected ultra-faint dwarf galaxies suitable for testing dark matter density profiles with wide-binary statistics.
}
\label{tab:dSphs_params}
  \small
  \begin{tabular}{ccccccccccccc}
  \hline
  Galaxy & RA [deg] & DEC [deg] & $R_{1}$ [$^\circ$] & $R_{i}$ [kpc] & $M_{V}$ [mag] & $\epsilon$ & Distance [kpc] & PA [$^\circ$] \\
  \hline
  Segue\,I & 151.7504 & 16.0769 & 3.93 & 29.00 & $-$1.30 & 0.32 & 23.00 & 75.00 \\
  Triangulum II & 33.3225 & 36.1719 & 2.50 & 32.00 & $-$4.20 & 0.30 & 28.40 & 73.00 \\
  Hydrus I & 37.3890 & $-$79.3089 & 6.60 & 53.00 & $-$4.71 & 0.20 & 27.60 & 97.00 \\
  Ursa Major II & 132.8726 & 63.1335 & 13.90 & 149.00 & $-$4.25 & 0.55 & 32.00 & $-$76.00 \\
  Coma Berenices & 186.7454 & 23.9069 & 5.67 & 77.00 & $-$4.38 & 0.37 & 44.00 & $-$58.00 \\
  \hline
  \end{tabular}
  
  \vspace{0.2cm}
  \noindent\footnotesize Table parameters are based on data from \citet{Tau2023}.
\end{table*}

%{\magenta JP: I would comment here that the presence of DM subhaloes predicted by CDM may also affect the survival of wide binaries 
%https://arxiv.org/abs/1005.5388
%and thus our constraints on the cusp/core profile.}
Additionally, the presence of dark matter subhaloes, as predicted by CDM models, may also influence the survival of wide binaries, as they can introduce dynamical interactions that could alter the binary population. This would impact the constraints on the cusp/core dark matter profile. \citet{Penarrubia2010} showed that such subhaloes could lead to the disruption or survival of wide binaries, and we plan to integrate this effect into future iterations of our model to improve the accuracy of the resulting dark matter profile constraints.

More fundamentally, it is important to emphasize that our attempt to distinguish between cored and cuspy dark-matter density profiles using wide binaries is inherently model dependent. Even if CSST observations were to yield a statistically significant detection of wide-binary populations in ultra-faint dwarf galaxies, the dynamical interpretation of such signals necessarily relies on assumptions about the formation channels and initial conditions of these systems. In practice, the present-day wide-binary population reflects the combined outcome of both formation and subsequent dynamical disruption, and these two processes cannot be treated independently.

In this work, we have adopted the survival prescriptions of
P16 as illustrative scenarios to assess the
observational sensitivity of CSST, rather than as definitive models of wide-binary evolution in UFDs. Different formation pathways or initial conditions could, in principle, produce wide-binary populations whose response to the same underlying dark-matter potential differs from those assumed here. As a result, any constraints on dark-matter core or cusp derived from wide-binary statistics should be interpreted within the context of the adopted formation and survival models. Future progress will require joint modeling of wide-binary formation and disruption, ideally informed by independent observational constraints, in order to fully exploit wide binaries as robust dynamical probes of dark matter in ultra-faint dwarf galaxies.
As shown in Figure \ref{fig:fb_recovery}, the direct 2PCF recovery of the binary fraction exhibits a systematic underestimation for shallower limiting magnitudes ($g_{\rm lim}\lesssim 26.5$) due to photometric incompleteness. While our deep-field assumption ($g\sim27.5$) mitigates this bias, future applications to shallower surveys or the outskirts of UFDs may benefit from a forward-modeling approach. Instead of fitting the 2PCF of the raw data, one can construct a generative model that includes the intrinsic binary fraction, the survival function $f_s(s,t)$, the CSST completeness and resolution limits, and then sample the posterior probability via a likelihood function. Such a forward model would naturally account for selection effects and could recover an unbiased $f_b$ even at moderate depth, at the cost of increased computational complexity. We defer this refinement to future work.

\subsection{Connection to Observational Strategy}

The methodological approach presented here—constraining dark matter profiles via the spatial analysis of wide binaries—has profound and direct implications for the observational strategy of UFDs using the CSST.This technique bypasses the demanding requirement of measuring individual, highly precise stellar proper motions or radial velocities for faint UFD members—a task often limited by the large distances and low luminosity of these targets. Instead, our method capitalizes on the CSST's exceptional deep imaging and wide-field capabilities. Specifically, the instrument's limiting magnitude of $g \approx 26.3$ (in $5\sigma$ detection) across a $1.1~\mathrm{deg}^2$ field of view, coupled with its high angular resolution ($\text{FWHM} < 0.15''$), is ideally suited to construct high-fidelity photometric catalogs of resolved stellar populations within UFDs.

Once a robust member catalog is established, the core observational task transitions from individual kinematic measurement to statistical spatial analysis via the 2PCF. The measured two-point correlation function quantifies the degree of spatial clustering of resolved stellar pairs, yielding a statistical measure of the binary fraction. Critically, because the survival probability and disruption rate of soft wide binaries are sensitive to the local tidal field and the DM density profile, the $2\text{PCF}$-derived distribution places direct statistical constraints on the DM structure within the UFD.

However,this statistical approach must be framed carefully. While the detection of wide binaries and the modeling of their survival can inform us on (i) the amount and distribution of DM in UFDs and potentially (ii) put a constraint on the number of low-mass, dark substructures within these galaxies, we must be cautious about making definitive statements regarding the fundamental nature of DM. Specifically, the analysis is intrinsically subject to a degeneracy between the wide binary's formation time/mechanism and the inferred DM parameters. Since the disruption of wide binaries is a time-dependent process, the uncertainty in their initial properties directly translates into uncertainty in the derived DM density profile. Therefore, the results derived from the 2PCF analysis must be rigorously compared against simulations adopting different DM models and accounting for a range of formation scenarios.

\subsection{Degeneracies from dark matter subhalos and stellar clusters}
Throughout this work, we have followed the framework of P16, which assumes a smooth DM potential. However, in realistic $\Lambda$ CDM and many alternative DM models, the halo contains a population of DM subhalos. These subhalos can perturb wide binaries through impulsive gravitational encounters, injecting energy and potentially dissolving binaries in a manner that may partially mimic or enhance the disruption expected from a cuspy central density profile. This introduces a degeneracy between the effects of the global halo shape and those of the substructure population.

The degeneracy is challenging to disentangle because cuspy halos already produce stronger tidal fields that suppress wide binary survival, while subhalo encounters add stochastic, spatially variable heating that can blur the distinctive signatures of the global profile \citep{Webb2018}. Conversely, in cored halos where tidal fields are weaker, wide binaries are expected to survive more readily, but the presence of even a moderate abundance of subhalos could artificially reduce the binary fraction, leading to a false inference of a cusp. Thus, without careful modeling, an observed binary fraction alone cannot uniquely distinguish between a cusp with few subhalos and a core with many subhalos.

Baryonic substructures further complicate the picture. Observations of the ultra-faint dwarf Eridanus II have revealed the presence of a central star cluster \citep{Crnjevic2016,Contenta2018}, suggesting that similar stellar systems may have existed historically in other UFDs, including Segue\,I. The formation, evolution, and eventual dissolution of such clusters can inject additional dynamical heating into the stellar component, enhancing the disruption rate of wide binaries beyond that due to DM alone. Moreover, recently disrupted clusters could produce small-scale spatial correlations that might be mistaken for binary signals. These effects are not accounted for in the smooth‑potential simulations of P16.

A potential concern is that recently disrupted star clusters could leave behind unrelaxed stellar overdensities that mimic wide binaries in the 2PCF. However, the phase mixing time in Segue\,I is sufficiently short to mitigate this issue. For a typical velocity dispersion $\sigma_v \sim 3.7 \text{ km s}^{-1}$ and half-light radius $r_h \sim 30 \text{ pc}$, the crossing time is $t_{\text{cross}} \equiv r_h / \sigma_v \sim 8 \text{ Myr}$. Phase mixing proceeds on a timescale comparable to or a few times the crossing time, typically $\sim 10\text{–}50$ Myr for Segue\,I.Since Segue\,I has likely remained relatively quiescent over most of its $\sim 10$ Gyr history, any historical star clusters likely dissolved billions of years ago. Consequently, their debris would now be fully phase-mixed and indistinguishable from the background. Only an extremely recent ($\lesssim 50 \text{ Myr}$) disruption could preserve coherent spatial substructure. The probability of catching a star cluster in the exact act of dissolution within the last $\sim 0.5\%$ of the galaxy's lifetime is extremely low. Furthermore, even in such a rare event, the macro-scale spatial correlation of a tidal stream is expected to fundamentally differ from the small-scale binary survival signal, given that wide binaries probe separations below $\sim 1,\mathrm{pc}$—much smaller than the typical size of the galaxy ($\sim 10^2,\mathrm{pc}$). Hence, contamination from disrupted clusters is unlikely to significantly bias our 2PCF analysis.

Taken together, both DM subhalos and baryonic star clusters are expected to contribute to the dynamical evolution of wide binaries. Disentangling their respective influences from the imprint of the global DM density profile is a non‑trivial task. It requires forward modeling that simultaneously includes: (i) a realistic distribution of DM subhalos, (ii) a model for the formation, evolution, and dissolution of star clusters in UFDs, and (iii) a treatment of wide binary formation and subsequent dynamical evolution. While such a comprehensive approach is beyond the scope of the present work, we emphasize that any future attempt to use wide binaries as precise probes of the cusp‑core structure must address these degeneracies. Our current results should be viewed as an assessment of CSST’s observational sensitivity under idealized smooth‑potential assumptions, providing a baseline upon which more realistic substructure models can be built.

The degeneracies discussed above—between the global core or cusp profile, DM substructure, and baryonic cluster remnants—cannot be broken by spatial clustering alone. A promising avenue is to combine the 2PCF forward model with likelihoods from stellar kinematics. Existing multi-epoch spectroscopic data for Segue\,I \citep{Simon2011} provide constraints on the velocity dispersion and, after accounting for binary contamination, on the underlying dark matter potential. A joint likelihood $\mathcal{L}_{\rm total} = \mathcal{L}_{\rm 2PCF}(f_b,\mathrm{profile})\times \mathcal{L}_{\rm kin}(\sigma_{v},\mathrm{profile})$ would allow simultaneous inference of the binary population and the gravitational potential, breaking the degeneracy between a cusp with few subhalos and a core with abundant substructure. While building such a combined model is beyond the scope of this paper, we note that the CSST's proper motion measurements \citep{nie2025} will further tighten kinematic constraints, enabling a more robust discrimination in future work.

\subsection{Promising Candidate UFDs for CSST Observations}

Recent studies highlighting the limitations of wide binary statistics in constraining dark matter profiles  are primarily instructive for target selection \citep{Shariat2025}, rather than indicative of a fundamental flaw in the 2PCF method. The analysis of Boötes I reveals that its large distance ($\sim 66$ kpc) severely limits angular resolution and contrast sensitivity, amplifying contamination from chance alignments at the critical separations needed to probe the inner gravitational potential. Conversely, while closer, Reticulum II suffers from an intrinsically low binary fraction of $f_{\rm b} \sim 0.007$ at $s > 3000$ au \citep{Safarzadeh2022}, resulting in a 2PCF signal too weak for robust dynamical discrimination. Thus, the key challenge lies not in the method itself, but in applying it to suboptimal systems. The 2PCF approach remains theoretically potent for nearer UFDs that also harbour a higher intrinsic binary fraction, where the signal-to-noise and resolving power are sufficient to disentangle dynamical signatures from systematic uncertainties.

From an observational perspective, extending this approach to upcoming wide-field surveys such as CSST or LSST is most promising for several nearby and dark-matter-dominated UFDs. Our analysis identifies a set of ideal candidates based on proximity and existing observational constraints.

Among them, Segue\,I serves as the benchmark system due to its extensive prior study. Triangulum\,II, located at a similar distance and exhibiting an extremely high mass-to-light ratio, would be an ideal target if its dynamical equilibrium is confirmed. Reticulum\,II at $\sim 30$~kpc, possess compact half-light radii and rich spectroscopic membership samples, which will enable effective foreground suppression and robust member selection. Ursa Major\,II and Coma Berenices, although slightly brighter and partially tidally disturbed, remain suitable for methodological validation after excluding their perturbed regions due to their proximity ($D \approx 30-45$~kpc).

In summary, Segue\,I, Triangulum\,II, and Hydrus\,I represent the most promising candidates for constraining dark matter density profiles via the 2PCF technique, while Ursa Major\,II and Coma Berenices can serve as boundary cases to test the robustness of the method. Their proximity and extreme dark-matter dominance suggest that, with forthcoming deep surveys, the angular separation statistics of wide binaries will likely reach the statistical significance required to discriminate between cusped and cored halo models.

\section{Conclusions}
\label{sec:6}
In this work, we have investigated the observational potential of the CSST to detect and statistically characterize wide-binary populations in UFDs, using Segue\,I as a benchmark system. Rather than attempting to directly infer the dark-matter density profile of these systems, our goal has been to assess the feasibility and sensitivity of a 2PCF–based approach for recovering wide-binary signatures under realistic observational conditions. The principal conclusions of our study can be summarized as follows:

\begin{enumerate}
    \item
 The CSST, with its deep imaging capabilities ($g \sim 27.5$ mag, and $\sim 90\%$ completeness), will enable the robust recovery of the intrinsic binary fraction in UFDs. This is a fundamental prerequisite for the application of our method.

    \item  The small-scale ($\sim 0.3-1$ pc) amplitude of the 2PCF is highly sensitive to the underlying gravitational potential. We quantitatively show that cusped dark matter halos ($\gamma \sim 1$) produce a systematically suppressed 2PCF signal compared to cored halos ($\gamma \sim 0$), providing a clear observational signature to distinguish between these competing models.
    \item Our detectability forecasts demonstrate that CSST can robustly recover wide-binary fractions at the percent level in nearby ultra-faint dwarfs such as Segue\,I, while for more distant systems ($\sim 30$ kpc) the required binary fractions rise to $\sim 5-10$ percent, providing a quantitative framework for interpreting future observations of stellar multiplicity in the lowest-mass galaxies.
    \item We find that the cumulative SNR depends strongly on total star number $N_{\text{star}}$ and initial binary fraction $f_b'$, indicating that CSST possesses the statistical sensitivity to discriminate between dark matter cusp and core profiles in nearby UFDs ($D \lesssim 40$ kpc) if $N_{\text{star}} \gtrsim 6000$ and $f_{b} \gtrsim 0.1$. We note, however, that the robustness of these dynamical constraints is inherently tied to the assumed initial distribution of wide binaries. Given the uncertainties in wide-binary formation mechanisms, these thresholds should be interpreted as model-dependent forecasts.

    \item
 More importantly, our 2PCF-based approach is its insensitive to the mass-ratios of binaries, making it robust against uncertainties in binary population models. Furthermore, it circumvents the long-standing systematic bias of unresolved binaries that plagues traditional kinematic studies based on stellar velocity dispersions.

    \item We identify a set of promising UFDs for applying this technique with upcoming surveys. Systems like Segue\,I, Triangulum~II and Hydrus~I are ideal primary targets due to their proximity and extreme dark matter dominance. Others, such as Ursa~Major~II and Coma~Berenices, serve as valuable boundary cases for testing the method's robustness against tidal influences.
\end{enumerate}

In summary, our investigation demonstrates that the CSST will provide the observational depth and statistical power necessary to measure the wide-binary population in ultra-faint dwarfs. The resulting two-point correlation function statistics for stellar pairs offer a promising pathway to test theoretical predictions linking wide-binary survival to the underlying gravitational potential.

Specifically, applying this methodology to a cohort of Local Group UFDs with CSST is expected to yield the first systematic, empirical measurements of binary fractions in these extreme environments. These measurements will provide a crucial observational benchmark against which the predictions of different dark matter halo models can be compared. Our forecasts demonstrate that, under the assumption that the predicted differences in wide-binary survival manifest on angular scales accessible to CSST, the 2PCF provides a statistically sensitive method to recover such signals from the projected stellar distribution, thereby informing the theoretical understanding of dark matter on sub-galactic scales and the dynamical mechanisms behind the formation of wide binaries.

{\it Acknowledgements.}
The authors thank Feng Wang, Chao Liu, and Xiaoting Fu for the helpful discussions. H.J.T. thanks the support from the NSFC grant (No. 12373033) and the Key Project of Zhejiang Provincial Natural Science Foundation (No. ZCLZ25A0301). B.Y. thanks the support from the NSFC International (Regional) Cooperation and Exchange Project (No. 12361141814).
\section*{Data availability}
The data underlying this article will be shared on reasonable request to the corresponding author.

\bibliographystyle{mnras}
\bibliography{reference} % if your bibtex file is called example.bib

% Alternatively you could enter them by hand, like this:
% This method is tedious and prone to error if you have lots of references
%\begin{thebibliography}{99}
%\bibitem[\protect\citeauthoryear{Author}{2012}]{Author2012}
%Author A.~N., 2013, Journal of Improbable Astronomy, 1, 1
%\bibitem[\protect\citeauthoryear{Others}{2013}]{Others2013}
%Others S., 2012, Journal of Interesting Stuff, 17, 198
%\end{thebibliography}

%%%%%%%%%%%%%%%%%%%%%%%%%%%%%%%%%%%%%%%%%%%%%%%%%%

%%%%%%%%%%%%%%%%% APPENDICES %%%%%%%%%%%%%%%%%%%%%

%\appendix

%\section{Some extra material}

%If you want to present additional material which would interrupt the flow of the main paper,
%it can be placed in an Appendix which appears after the list of references.

%%%%%%%%%%%%%%%%%%%%%%%%%%%%%%%%%%%%%%%%%%%%%%%%%%

% Don't change these lines
\bsp	% typesetting comment
\label{lastpage}
\end{document}